\documentclass[preprint,floatfix] {revtex4} 
 
\usepackage{graphicx}
\usepackage{subfigure}
\begin{document}

\title{Ro-vibrational studies of diatomic molecules in a shifted Deng-Fan oscillator potential}
\author{Amlan K. Roy}
\altaffiliation{Email: akroy@iiserkol.ac.in, akroy6k@gmail.com \\
This article is dedicated to my kind-hearted father Sj.~Bisweswar Roy, on the occasion of his 70th birthday. It is because of his untiring 
effort and constant inspiration that I could pursue research.}
\affiliation{Division of Chemical Sciences,   
Indian Institute of Science Education and Research (IISER)-Kolkata, 
Mohanpur Campus, P. O. BCKV Campus Main Office, Nadia, 741252, WB, India.}

\begin{abstract}
Bound-state spectra of shifted Deng-Fan oscillator potential are studied by means of a generalized pseudospectral method. Very accurate 
results are obtained for \emph{both low as well as high} states by a non-uniform optimal discretization of the radial Schr\"odinger equation. 
Excellent agreement with literature data is observed in \emph{both $s$-wave and rotational} states. Detailed variation of energies with
respect to potential parameters is discussed. Application is made to the ro-vibrational levels of four representative diatomic molecules 
(H$_2$, LiH, HCl, CO). Nine states having $\{n,\ell\} =0,1,2$ are calculated with good accuracy along with 15 other higher states for each 
of these molecules. Variation of energies with respect to state indices $n$, $\ell$ show \emph{behavior} similar to that in the Morse 
potential. Many new states are reported here for the first time. In short, a simple, accurate and efficient method is presented for this and 
other similar potentials in molecular physics. 

\end{abstract}
\maketitle

\section{Introduction}
Interest in empirical potential energy functions of diatomic molecules has remained unabated ever since the inception of the celebrated
Morse potential \cite{morse29}, proposed nearly 85 years ago. This three-parameter, exponentially varying function has 
inspired numerous works in the direction of constructing a universal energy-distance relationship, having varied number of parameters. 
In recent years there has been an upsurge of interest in this direction. Thus, an enormous number of publications exist in the 
literature spanning over several decades. Some of the most prominent ones include Morse \cite{morse29,dong02a,nasser07}, Rydberg 
\cite{rydberg31}, P\"oschl-Teller \cite{poschl33,dong02}, Manning-Rosen potential in 3D \cite{manning33,qiang07,qiang09}, in higher dimension 
\cite{gu11}, in relativistic domain \cite{wei08,wei10} and scattering states \cite{wei08a}, Rosen-Morse \cite{rosen32}, Wood-Saxon 
\cite{ikhdair10}, pseudoharmonic potential \cite{dong03,sever07} in 2D \cite{dong02b}, Kratzer \cite{kratzer20}, Hulth\'en \cite{hulthen42}, 
hyperbolic \cite{dong08}, Linnett \cite{linnett40}, Lippincott \cite{lippincott53}, Tietz \cite{tietz63}, Schl\"oberg \cite{schloberg86}, Zavitsas 
\cite{zavitsas91}, Hajigeorgiou \cite{haji10}, along with many other variants of these and numerous others. 

In this work, we focus on the important Deng-Fan (DF) potential \cite{deng57} for diatomic molecules, proposed little more than half a 
century ago, but attracting much interest lately,
\begin{equation}
V(r)=D_e \left( 1- \frac{b} {e^{a r}-1} \right)^2, \ \ \ \ b=e^{a r_e} -1, \ \ \ \  r \in (0,\infty).
\end{equation}
The three positive parameters $D_e, r_e, a$ denote dissociation energy, equilibrium internuclear distance and the radius of potential
well respectively. It shows qualitatively correct asymptotic behavior as internuclear distance tends towards zero and infinity. Because of the qualitative 
similarity with Morse potential, this is also often termed as Generalized Morse potential \cite{mesa98,rong03}. Besides, this is also 
related to the well-known Manning-Rosen potential \cite{manning33} and has found interesting applications in molecular spectroscopy as well as 
electronic transitions. Employing this potential, transition frequencies and intensities of overtones of X-H stretching vibrations in small 
molecules were calculated and compared with Morse potential \cite{rong03}. 

As with many other potentials of interest in physical, chemical systems, \emph{exact} analytical solution of this potential for 
\emph{arbitrary} quantum states has not been found as yet. Therefore, 
several theoretical attempts have been made to approximate the nature of solutions of this potential in relativistic and non-relativistic 
domain. For example, the exact solvability problem was discussed by means of an $SO(2,2)$ symmetry algebra \cite{mesa98}. Later, 
eigenvalues, eigenfunctions for $\ell=0$ were investigated by the $SO (2,1)$ algebraic method \cite{codriansky99}. About a decade later, 
approximate analytical solutions of Schr\"odinger equation with rotating DF potential for arbitrary $n, \ell$ states were presented in 
terms of the generalized hypergeometric functions $_2F_1(a,b;c;z)$ \cite{dong08a}. In another development, an improved approximation scheme 
was used for the centrifugal term, along with a super-symmetric shape invariance approach \cite{zhang11}. Then, using a 
super-symmetric shape invariance formalism, approximate analytic solution of the Dirac equation with DF potential has been given 
\cite{zhang09}. Bound state solutions of the Klein-Gordon equation with rotating DF potential has been presented as well lately for 
spinless particle \cite{dong11}. Analytic solutions of the Klein-Gordon and Dirac equation 
in a rotating DF potential is given by a Pekeris approximation of the centrifugal term and a Nikiforov-Uvarov method \cite{oluwadare12}. Moreover, 
it has been recently demonstrated that, for a set of 16 diatomic molecules Manning-Rosen, Schl\"oberg and DF potentials perform very closely 
to the traditional Morse potential (and not showing any significant improvement over it) in terms of spectroscopic parameters, even though 
the latter follows qualitatively correct asymptotic behavior \cite{wang12}.
 
Recently, the DF potential, shifted by the dissociation energy, has been used for molecules \cite{hamzavi13}, 
\begin{equation}
V(r)=D_e \left( 1- \frac{b} {e^{a r}-1} \right)^2 -D_e
    =D_e \left[ \frac{b^2}{(e^{ar}-1)^2} - \frac{2b}{e^{ar}-1} \right] , \ \ \ \ b=e^{a r_e} -1.
\end{equation}
The shifted DF (sDF) potential in Eq.~(2) resembles the behavior of Morse potential for large $r$ regions ($r \approx r_e, r > r_e$), but 
differs at $r \approx 0$. Moreover, deep DF and sDF potentials ($D_e \gg 1$), can be approximated by harmonic oscillator 
in the $r \approx r_e$ region \cite{mesa98,rong03}. 

The purpose of this work is to investigate the bound-state spectra of DF-type potential for \emph{arbitrary} quantum numbers $n, \ell$, 
as well as for both \emph{low and high} states. Although some decent results are available \cite{dong08a, zhang11, hamzavi13, oyewumi13}, 
there is a need of 
systematic analysis in terms of accurate eigenvalues, eigenvalues and other relevant quantities, especially for the situations
mentioned above. For this we employ the generalized pseudospectral method (GPS), which has been very successful for a number
of physically important quantum systems, such as, spiked harmonic oscillator, rational, Hulth\'en, Yukawa, logarithmic, power-law, Morse 
potential as well as static and dynamic studies in molecules and atoms (including Rydberg states), etc. \cite{roy04, roy05, roy05a, roy08, 
roy13, roy13a}. Thus, at first, a detailed study is presented on the bound states of DF potential covering 36 states 
corresponding to vibrational quantum number 
up to $n=7$. A thorough variation of energies with respect to the parameters $a, r_e$ is monitored. Then it is extended for 
ro-vibrational levels (belonging to both low and high $n,\ell$) within sDF potential, for four diatomic molecules, namely, H$_2$, LiH, HCl 
and CO. This will also broaden the range of applicability and feasibility of GPS method for a larger class of complicated molecular 
potential functions. Comparison with existing literature data is made wherever possible. General qualitative behavior of ro-vibrational
energies obtained from this potential is also briefly contrasted with those from the familiar Morse potential. The article is organized as 
follows: A brief account of the employed GPS method is provided in Section II. Section III gives results and discussion, while a few concluding 
remarks are noted in Section V. 

\section{GPS method for DF potential}
In this section, we briefly outline the GPS formalism for solving the non-relativistic Schr\"odinger equation of a Hamiltonian containing 
a DF potential. Only the essential steps are given; relevant details may be found in previous works 
\cite{roy04, roy05, roy05a, roy08, roy13, roy13a} and the references therein). Atomic units are employed 
throughout the article, unless otherwise mentioned. 

The desired radial Schr\"odinger equation to be solved, can be written in following form, 
\begin{equation}
\left[-\frac{1}{2} \ \frac{\mathrm{d^2}}{\mathrm{d}r^2} + \frac{\ell (\ell+1)} {2r^2}
+v(r) \right] \psi_{n,\ell}(r) = E_{n,\ell}\ \psi_{n,\ell}(r)
\end{equation}
where $v(r)$ is DF or sDF potential, given in Eq.~(1) or (2). Here $n$, $\ell$ signify the usual radial and angular momentum quantum numbers 
respectively. 

The characteristic feature of GPS method lies in the fact that it facilitates the use of a \emph{non-uniform, optimal} spatial discretization
leading to a coarser mesh at larger $r$ and denser mesh at smaller $r$, while maintaining similar accuracy at both these regions. Thus it 
enables one to work with a significantly smaller number of grid points yet providing accurate results quite efficiently. This is in sharp
contrast to the usual finite-difference schemes, which require considerably larger grid points, presumably because of their uniform 
discretization nature. 

At first, a function $f(x)$ defined in the interval $x \in [-1,1]$, is approximated by an N-th order polynomial $f_N(x)$, as given below,
\begin{equation}
f(x) \cong f_N(x) = \sum_{j=0}^{N} f(x_j)\ g_j(x),
\end{equation}
which guarantees that the approximation is \emph {exact} at the \emph {collocation points} $x_j$, i.e., $f_N(x_j) = f(x_j).$
In the Legendre pseudospectral scheme used here, $x_0=-1$, $x_N=1$, while $x_j$ $(j=1,\ldots,N-1)$ are obtained from the roots of the first 
derivative of Legendre polynomial $P_N(x)$ with respect to $x$, i.e., $P'_N(x_j) = 0.$ The cardinal functions, $g_j(x)$ in Eq.~(4) are given by,
\begin{equation}
g_j(x) = -\frac{1}{N(N+1)P_N(x_j)}\ \  \frac{(1-x^2)\ P'_N(x)}{x-x_j},
\end{equation}
satisfying the unique property that $g_j(x_{j'}) = \delta_{j'j}$. In the next step, the semi-infinite domain $r \in [0, \infty]$ is mapped onto 
a finite domain $x \in [-1,1]$ by a transformation of the type $r=r(x)$. Now an algebraic nonlinear mapping of the following type can 
be introduced, 
\begin{equation}
r=r(x)=L\ \ \frac{1+x}{1-x+\alpha},
\end{equation}
with L, $\alpha= \frac{2L}{r_{max}}$ being two mapping parameters. Then applying a symmetrization procedure, one obtains, after some 
straightforward algebra, the following transformed Hamiltonian, 
\begin{equation}
\hat{H}(x)= -\frac{1}{2} \ \frac{1}{r'(x)}\ \frac{d^2}{dx^2} \ \frac{1}{r'(x)}
+ v(r(x))+v_m(x),
\end{equation}
where $v_m(x)$ is given by,
\begin{equation}
v_m(x)=\frac {3(r'')^2-2r'''r'}{8(r')^4}.
\end{equation}
The advantage is that now one deals with a \emph {symmetric} matrix eigenvalue problem, which can be easily solved by standard available 
routines to yield both eigenvalues and eigenfunctions very accurately and efficiently. 
 
\begingroup
\squeezetable
\begin{table}
\caption {\label{tab:table1} Comparison of calculated eigenvalues (a.u.) of DF potential for selected values of $a$. Parameters $r_e$ and 
$D_e=15$ are fixed at 0.4 and 15. PR signifies Present Result.} 
\begin{ruledtabular}
\begin{tabular}{ccccccc}
State & $a$ & \multicolumn{2}{c}{Energy} &  $a$ &  \multicolumn{2}{c}{Energy} \\ 
\cline{3-7} 
    &      & PR            & Literature                           &      &  PR            & Literature \\   \hline
$1s$  & 0.05 &  5.526780278  &                                      & 0.15 &  5.660016068   &            \\
    & 0.25$^{\dagger}$ &  5.792991281  &                                      & 0.35 &  5.925706889   &            \\
    & 0.45 &  6.058162830  &                                      & 0.55 &  6.190358005   &            \\
$2s$  & 0.05 & 10.024820283  &                                      & 0.15 & 10.324010665   &            \\
    & 0.25$^{\dagger}$ & 10.615568648  &                                      & 0.35 & 10.899365004   &            \\
    & 0.45 & 11.175264491  &                                      & 0.55 & 11.443125859   &            \\
$2p$  & 0.05 &  7.860804466  & 7.8606\footnotemark[1],7.86080\footnotemark[2],7.860804467\footnotemark[3],7.8628\footnotemark[4] 
    & 0.15 &  8.045097775  & 8.04322\footnotemark[1],8.04510\footnotemark[2],8.045099635\footnotemark[3],8.04724\footnotemark[4]  \\
    & 0.25 &  8.226613566  & 8.22142\footnotemark[1],8.22663\footnotemark[2],8.226628516\footnotemark[3],8.22892\footnotemark[4] 
    & 0.35 &  8.405438375  &                                                                                                      \\
    & 0.45 &  8.581651488  &                                      & 0.55 &  8.755325874   &            \\
$3s$  & 0.05 & 12.003203832  &                                      & 0.15 & 12.381945370   &            \\
    & 0.25$^{\dagger}$ & 12.737133331  &                                      & 0.35 & 13.068350233   &            \\
    & 0.45 & 13.375161765  &                                      & 0.55 & 13.657116713   &            \\
$3p$  & 0.05 & 10.997762943  & 10.9976\footnotemark[1],10.9978\footnotemark[2],10.99776302\footnotemark[3],10.9998\footnotemark[4] 
    & 0.15 & 11.324240817  & 11.3224\footnotemark[1],11.3242\footnotemark[2],11.32424848\footnotemark[3],11.32647\footnotemark[4]   \\
    & 0.25 & 11.638278167  & 11.6331\footnotemark[1],11.6383\footnotemark[2],11.63833602\footnotemark[3],11.64068\footnotemark[4] 
    & 0.35 & 11.939885166  &                                                                                                      \\
    & 0.45 & 12.229041783  &                                      & 0.55 & 12.505700581   &            \\
$3d$  & 0.05 & 10.215980103  & 10.2154\footnotemark[1],10.21598\footnotemark[2],10.21598019\footnotemark[3],10.21651\footnotemark[4] 
    & 0.15 & 10.489341948  & 10.4837\footnotemark[1],10.48935\footnotemark[2],10.48935369\footnotemark[3],10.48992\footnotemark[4]   \\
    & 0.25 & 10.755814653  & 10.7403\footnotemark[1],10.75591\footnotemark[2],10.75590446\footnotemark[3],10.74645\footnotemark[4] 
    & 0.35 & 11.015610519  &                                                                                                      \\
    & 0.45 & 11.268908225  &                                      & 0.55 & 11.515858603   &            \\
$4s$  & 0.05 & 13.043507261  &                                      & 0.15 & 13.455615519   &            \\
    & 0.25$^{\dagger}$ & 13.821048531  &                                      & 0.35 & 14.138978827   &            \\
$4p$  & 0.05 & 12.497602157  & 12.4974\footnotemark[1],12.4976\footnotemark[2],12.49760240\footnotemark[3],12.4992\footnotemark[4] 
    & 0.15 & 12.888327591  & 12.8865\footnotemark[1],12.88835\footnotemark[2],12.88834790\footnotemark[3],12.8901\footnotemark[4]   \\
    & 0.25 & 13.248318043  & 13.2433\footnotemark[1],13.24847\footnotemark[2],13.24846979\footnotemark[3],13.2501\footnotemark[4] 
    & 0.35 & 13.577277786  &                                                                                                      \\
$4d$  & 0.05 & 12.098289743  & 12.0977\footnotemark[1],12.09829\footnotemark[2],12.09829019\footnotemark[3],12.0989\footnotemark[4] 
    & 0.15 & 12.466379229  & 12.4608\footnotemark[1],12.46642\footnotemark[2],12.46641867\footnotemark[3],12.46715\footnotemark[4]   \\
    & 0.25 & 12.813205240  &                                      & 0.35 & 13.138823092  &             \\
$4f$  & 0.05 & 11.820785582  & 11.8195\footnotemark[1],11.82079\footnotemark[2],11.82078608\footnotemark[3],11.8209\footnotemark[4] 
    & 0.15 & 12.171646579  & 12.1604\footnotemark[1],12.17170\footnotemark[2],12.17169520\footnotemark[3],12.1718\footnotemark[4]   \\
    & 0.25 & 12.508800391  &                                      & 0.35 & 12.832495054  &             \\
$5s$  & 0.1  & 13.874627559  &                                      & 0.25$^{\dagger}$ & 14.416949441  &             \\
$5p$  & 0.1  & 13.542133643  & 13.5413\footnotemark[1],13.54214\footnotemark[2],13.54214240\footnotemark[3],13.5434\footnotemark[4] 
    & 0.25 & 14.101049462  &             \\
$5d$  & 0.1  & 13.306777642  & 13.3043\footnotemark[1],13.30680\footnotemark[2],13.30679659\footnotemark[3],13.3075\footnotemark[4] 
    & 0.25 & 13.868649955  &             \\
$5f$  & 0.1  & 13.147569396  & 13.1426\footnotemark[1],13.14760\footnotemark[2],13.14759709\footnotemark[3],13.1478\footnotemark[4] 
    & 0.25 & 13.712818743  &             \\
$5g$  & 0.1  & 13.037943909  & 13.0296\footnotemark[1],13.03798\footnotemark[2],13.03797516\footnotemark[3],13.0379\footnotemark[4] 
    & 0.25 & 13.611549224  &             \\
$6s$  & 0.1  & 14.262988907  &                                      & 0.25$^{\dagger}$ & 14.749124380  &             \\
$6p$  & 0.1  & 14.052071899  & 14.0513\footnotemark[1],14.05209\footnotemark[2],14.05208850\footnotemark[3],14.0530\footnotemark[4] 
    & 0.25 & 14.575579749  &             \\
$6d$  & 0.1  & 13.907009810  & 13.9045\footnotemark[1],13.90705\footnotemark[2],13.90704815\footnotemark[3],13.9075\footnotemark[4] 
    & 0.25 & 14.449406270  &             \\
$6f$  & 0.1  & 13.811128402  & 13.8062\footnotemark[1],13.81119\footnotemark[2],13.81118932\footnotemark[3],13.8113\footnotemark[4] 
    & 0.25 & 14.368521327  &             \\
$6g$  & 0.1  & 13.746532778  & 13.7383\footnotemark[1],13.74661\footnotemark[2],13.74661179\footnotemark[3],13.7466\footnotemark[4] 
    & 0.25 & 14.320719404  &             \\
$6h$  & 0.1  & 13.701813086  &                                      & 0.25 & 14.296740289  &             \\
\end{tabular}
\end{ruledtabular}
\begin{tabbing}
$^{\mathrm{a}}${Ref.~\cite{dong08a}.}  \hspace{75pt} \= 
$^{\mathrm{b}}${Ref.~\cite{zhang11}.}  \hspace{75pt} \= 
$^{\mathrm{c}}${Ref.~\cite{oyewumi13}.}  \hspace{75pt} \= 
$^{\mathrm{d}}${Ref.~\cite{lucha99}}, as quoted in \cite{dong08a}.   \\
$^{\dagger}${See the Supplementary Material for results of these states, by an ``Anonymous Referee".}
\end{tabbing}
\end{table}
\endgroup

\section{Results and Discussion}
At first, we give our central result of bound-state energies $E_{n,\ell}$ of DF potential obtained from GPS method, for both non-rotational 
($\ell =0$) and rotational ($\ell \neq 0$) cases. For this, 21 energies belonging to the radial quantum number $n=0-5$ are reported, in a.u. 
Tables I and II correspond to $r_e=0.4$ and 0.8 respectively, while $D_e$ is kept fixed at 15 in both cases, in order to facilitate 
comparison with literature results. The potential parameter $a$ is varied from 0.05--0.45 to cover a broad range of interaction. A few 
approximate analytical and numerical results have been published recently for some of these states, which are quoted here appropriately.  
Note that in all the calculations in the present tables and also in the following, GPS mapping parameters $L=25, N=300$ were chosen, while the 
maximum radial distance $r_{max}$ needed to be adjusted for higher lying states. For lower states, generally a value of 500 a.u. was found 
to be necessarily sufficient; however for \emph{high-lying} states and \emph{large} radius ($a$) of the potential, larger values (up to even a 
few thousand a.u.) was required to capture the complicated nature of long-range tail in the wave functions. Similar situation was encountered for 
the Hulth\'en and Yukawa potentials in \emph{higher} states and for \emph{stronger} screening parameters \cite{roy05}. This is felt more so if 
high accuracy is desirable. However, the energies remained apparently completely insensitive to the variations in total number of radial points, 
as long as a decent number of collocation points were employed for sampling. Thus there is no computational overhead for this extension of the 
grid. This consistent set of parameters was adopted, after performing a series of calculations to reproduce the best existing energies in the 
literature. All our converged energies 
reported here are \emph{truncated} rather than rounded-off. No direct results are available for the $s$ ($\ell=0$) states for comparison. 
For the non-zero $\ell$ states, first systematic, good-quality approximate analytical energies were reported in \cite{dong08a}, which 
expressed them in terms of hypergeometric functions \cite{dong08a}. Approximate analytical energies from super-symmetric shape invariance 
formalism in conjunction with the wave function analysis \cite{zhang11} has produced slightly better eigenvalues. For the same parameter sets, 
approximate ro-vibrational states have also been reported through asymptotic iteration scheme along with a Pekeris-type scheme for the 
centrifugal term \cite{oyewumi13}. Additionally, eigenvalues are available from a MATHEMATICA implementation 
\cite{lucha99}, as quoted in \cite{dong08a}. The GPS eigenvalues show excellent agreement with all these results overall. As one goes to 
higher states, considerable difference in energies is noticed between those of \cite{zhang11} and \cite{oyewumi13}. Present energies tend to
differ from those of \cite{oyewumi13} for larger $a$ and higher states. We also note that all the six $\ell=0$ states were independently
obtained by an anonymous referee by using a Numerov-Cooley algorithm. These, given in the Supplementary Material, use a radial grid covering
0.0001 to 55.05 a.u., and match excellently with the current GPS results. 

\begingroup
\squeezetable
\begin{table}
\caption {\label{tab:table2} Comparison of calculated eigenvalues (a.u.) of DF potential for selected values of $a$. Parameters $r_e, D_e$ are 
fixed at 0.8 and 15. PR signifies Present Result.} 
\begin{ruledtabular}
\begin{tabular}{ccccccc}
State & $a$ & \multicolumn{2}{c}{Energy} &  $a$ &  \multicolumn{2}{c}{Energy} \\ 
\cline{3-7} 
    &      & PR            & Literature                           &      &  PR            & Literature \\   \hline
$1s$  & 0.05 &  3.123075639  &                                      & 0.15 &  3.260690978   &            \\
    & 0.25 &  3.399870299  &                                      & 0.35 &  3.540603794   &            \\
    & 0.45 &  3.682875356  &                                      & 0.55 &  3.826662669   &            \\
$2s$  & 0.05 &  6.938065055  &                                      & 0.15 &  7.301740816   &            \\
    & 0.25 &  7.662186480  &                                      & 0.35 &  8.019137561   &            \\
    & 0.45 &  8.372302228  &                                      & 0.55 &  8.721361792   &            \\
$2p$  & 0.05 &  4.140887222  & 4.14068\footnotemark[1],4.140887\footnotemark[2],4.140887237\footnotemark[3],4.14208\footnotemark[4] 
    & 0.15 &  4.297390050  & 4.29552\footnotemark[1],4.297393\footnotemark[2],4.297392964\footnotemark[3],4.2987\footnotemark[4]  \\
    & 0.25 &  4.453636191  & 4.44845\footnotemark[1],4.453660\footnotemark[2],4.453659003\footnotemark[3],4.4551\footnotemark[4] 
    & 0.35 &  4.609754444  &                                                                                                      \\
    & 0.45 &  4.765856216  &                                      & 0.55 &  4.922036867   &            \\
$3s$  & 0.05 &  9.236089799  &                                      & 0.15 &  9.757803393   &            \\
    & 0.25 & 10.261477247  &                                      & 0.35 & 10.746240649   &            \\
    & 0.45 & 11.211158603  &                                      & 0.55 & 11.655233084   &            \\
$3p$  & 0.05 &  7.532791457  & 7.53258\footnotemark[1],7.532792\footnotemark[2],7.532791535\footnotemark[3],7.5350\footnotemark[4] 
    & 0.15 &  7.915170747  &  7.9133\footnotemark[1],7.915179\footnotemark[2],7.915178421\footnotemark[3],7.9177\footnotemark[4]   \\
    & 0.25 &  8.291296319  & 8.28615\footnotemark[1],8.291354\footnotemark[2],8.291353518\footnotemark[3],8.2941\footnotemark[4] 
    & 0.35 &  8.661130740  &                                                                                                      \\
    & 0.45 &  9.024584285  &                                      & 0.55 &  9.381519299   &            \\
$3d$  & 0.05 &  5.739751067  & 5.73913\footnotemark[1],5.739751\footnotemark[2],5.739751150\footnotemark[3],5.7404\footnotemark[4] 
    & 0.15 &  5.950665807  & 5.94505\footnotemark[1],5.950678\footnotemark[2],5.950677430\footnotemark[3],5.9515\footnotemark[4]   \\
    & 0.25 &  6.157304825  & 6.14177\footnotemark[1],6.157395\footnotemark[2],6.157393368\footnotemark[3],6.1582\footnotemark[4] 
    & 0.35 &  6.360039366  &                                                                                                      \\
    & 0.45 &  6.559202848  &                                      & 0.55 &  6.755096503   &            \\
$4s$  & 0.05 & 10.725402154  &                                      & 0.15 & 11.351219388   &            \\
    & 0.25 & 11.934996174  &                                      & 0.35 & 12.474902845   &            \\
$4p$  & 0.05 &  9.613012874  & 9.6128\footnotemark[1],9.613013\footnotemark[2],9.613013061\footnotemark[3],9.6156\footnotemark[4] 
    & 0.15 & 10.148539652  & 10.1467\footnotemark[1],10.14856\footnotemark[2],10.14855549\footnotemark[3],10.1514\footnotemark[4]   \\
    & 0.25 & 10.661857334  & 10.6568\footnotemark[1],10.66197\footnotemark[2],10.66197323\footnotemark[3],10.665\footnotemark[4] 
    & 0.35 & 11.152379083  &                                                                                                      \\
$4d$  & 0.05 &  8.493343095  & 8.49272\footnotemark[1],8.493344\footnotemark[2],8.493343408\footnotemark[3],8.4948\footnotemark[4] 
    & 0.15 &  8.917778045  & 8.91218\footnotemark[1],8.917808\footnotemark[2],8.917806896\footnotemark[3],8.9194\footnotemark[4]   \\
    & 0.25 &  9.330059486  &                                      & 0.35 & 9.7305190720  &             \\
$4f$  & 0.05 &  7.434705351  & 7.43346\footnotemark[1],7.434706\footnotemark[2],7.434705654\footnotemark[3],7.4351\footnotemark[4] 
    & 0.15 & 7.735697652   & 7.72448\footnotemark[1],7.735732\footnotemark[2],7.735730867\footnotemark[3],7.7361\footnotemark[4]  \\
    & 0.25 &  8.027355594  &                                      & 0.35 &  8.310313463  &             \\
$5s$  & 0.1  & 12.098288273  &                                      & 0.25 & 13.049368752  &             \\
$5p$  & 0.1  & 11.302066518  & 11.3012\footnotemark[1],11.30207\footnotemark[2],11.30207233\footnotemark[3],11.3047\footnotemark[4] 
    & 0.25 & 12.200714709  &             \\
$5d$  & 0.1  & 10.520074121  & 10.5176\footnotemark[1],10.52009\footnotemark[2],10.52008576\footnotemark[3],10.5219\footnotemark[4] 
    & 0.25 & 11.332375565  &             \\
$5f$  & 0.1  &  9.796641911  & 9.79166\footnotemark[1],9.796658\footnotemark[2],9.796657408\footnotemark[3],9.7975\footnotemark[4] 
    & 0.25 & 10.503368829  &             \\
$5g$  & 0.1  &  9.152206082  & 9.14389\footnotemark[1],9.152223\footnotemark[2],9.152222313\footnotemark[3],9.1524\footnotemark[4] 
    & 0.25 &  9.747441273  &             \\
$6s$  & 0.1  & 12.846749917  &                                      & 0.25 & 13.803617463  &             \\
$6p$  & 0.1  & 12.279789391  & 12.279\footnotemark[1],12.27980\footnotemark[2],12.27979911\footnotemark[3],12.2822\footnotemark[4] 
    & 0.25 & 13.230098997  &             \\
$6d$  & 0.1  & 11.736417552  & 11.7339\footnotemark[1],11.73644\footnotemark[2],11.73643833\footnotemark[3],11.7383\footnotemark[4] 
    & 0.25 & 12.649721673  &             \\
$6f$  & 0.1  & 11.244784180  & 11.2398\footnotemark[1],11.24481\footnotemark[2],11.24481430\footnotemark[3],11.2459\footnotemark[4] 
    & 0.25 & 12.104277133  &             \\
$6g$  & 0.1  & 10.815295233  & 10.807\footnotemark[1],10.81533\footnotemark[2],10.81533124\footnotemark[3],10.8158\footnotemark[4] 
    & 0.25 & 11.615645551  &             \\
$6h$  & 0.1  & 10.446731008  &                                      & 0.25 & 11.190499481  &             \\
\end{tabular}
\end{ruledtabular}
\begin{tabbing}
$^{\mathrm{a}}${Ref.~\cite{dong08a}.}  \hspace{75pt} \= 
$^{\mathrm{b}}${Ref.~\cite{zhang11}.}  \hspace{75pt} \= 
$^{\mathrm{c}}${Ref.~\cite{oyewumi13}.}  \hspace{75pt} \= 
$^{\mathrm{d}}${Ref.~\cite{lucha99}}, as quoted in \cite{dong08a}.  
\end{tabbing}
\end{table}
\endgroup

Once the satisfactory performance for lower states is established, we now turn our focus on some select higher states of DF potential. To our 
knowledge, vibrational and rotational quantum numbers beyond 5 have not been considered before. Thus as a test of the validity and reliability of 
this approach, 15 states corresponding to $n=6,7$ (i.e., $7s, 7p, \cdots, 7i; 8s, 8p, \cdots, 8k$) are reported here in Table III, for the first 
time. Two values of $r_e=0.4$ and 0.8 are used, while $D_e$ remains constant at 15 in all cases. Discomfitures of some approximate (analytical 
as well as numerical) methods with higher states are well known and it is hoped that the present results would be helpful for future 
investigations. As in Table I, in this case also, some $\ell=0$ states ($7s, 8s, 9s$) were calculated for $a=0.1, r_e=0.4, D_e=15$ by the 
anonymous referee using Numerov-Cooley algorithm, and given in the Supplementary Material. Once again, these reproduce our results very nicely. 

\begingroup
\squeezetable
\begin{table}
\caption {\label{tab:table3}Calculated eigenvalues (a.u.) of DF potential for selected higher states ($n=$6,7) for $r_e=0.4$ and 0.8.
The parameters $D_e, a$ have been kept fixed at 15 and 0.1 respectively.} 
\begin{ruledtabular}
\begin{tabular}{ccllccll}
State & $a$ & \multicolumn{2}{c}{Energy}  & State    &  $a$ &  \multicolumn{2}{c}{Energy}  \\ 
\cline{3-4} \cline{7-8}
      &      &    $r_e=0.4$     &  $r_e=0.8$     &      &      &      $r_e=0.4$     &   $r_e=0.8$        \\  \hline
$7s$    & 0.1$^{\dagger}$  & 14.5181645271    & 13.3933271044  & $8s$   & 0.1$^{\dagger}$  & 14.6907030278      & 13.8011722762      \\
$7p$    & 0.1  & 14.3786954503    & 12.9782829487  & $8p$   & 0.1  & 14.5962052582      & 13.4909966267      \\
$7d$    & 0.1  & 14.2847492882    & 12.5879110006  & $8d$   & 0.1  & 14.5334902559      & 13.2034142273      \\
$7f$    & 0.1  & 14.2238628933    & 12.2407344479  & $8f$   & 0.1  & 14.4935808384      & 12.9510724239      \\
$7g$    & 0.1  & 14.1837577456    & 11.9420343928  & $8g$   & 0.1  & 14.4679688582      & 12.7366050653      \\
$7h$    & 0.1  & 14.1568264140    & 11.6891397269  & $8h$   & 0.1  & 14.4514673459      & 12.5570430884      \\
$7i$    & 0.1  & 14.1386698659    & 11.4761900800  & $8i$   & 0.1  & 14.4411087651      & 12.4074204733      \\
        &      &                  &                & $8k$   & 0.1  & 14.4351611618      & 12.2826683019      \\
\end{tabular}
\begin{tabbing}
$^{\dagger}${See the Supplementary Material for results of these states, by an ``Anonymous Referee".}
\end{tabbing}
\end{ruledtabular}
\end{table}
\endgroup

For further understanding, next we consider energy changes in DF potential with respect to the two parameters $a$ and $r_e$. In the lower (a) 
and upper (b) portions of left panel, variations of energy with respect to the parameter $a$ is shown at constant $r_e=0.4$, for $np \ (n=1-6)$ 
and $n'd \ (n'=2-7)$ series respectively. A large range of $a$ was allowed, although for convenience, the $n'd$ series, considers $a$ only up 
to 2.5.  The qualitative pattern of the plots in two series (a), (b) are very similar; showing a gradual increase with increase in $a$. The 
$n$ and $(n+1)$ levels remain well separated for the lowest $n$. However, the same becomes progressively smaller with increase in radial quantum 
number $n$. Likewise, in the lower (c) and upper (d) segments in the right panel, we monitor energy changes against $r_e$ (for a constant $a=0.1$), 
for the same two series of states, $np$ and $n'd$. Note that the range of $r_e$ is the same in (c), (d). Once again the $p$, $d$ series suggest 
quite similar energy behavior in (c), (d); initially the plots show a very sharp decline in energy, followed by a slow decrease and finally tend to 
assume a rather flat shape. In this occasion, however, for the value of vibrational quantum number studied, apparently the levels remain visibly 
well separated. 

\begin{figure}
\begin{minipage}[c]{0.40\textwidth}
\centering
\includegraphics[scale=0.38]{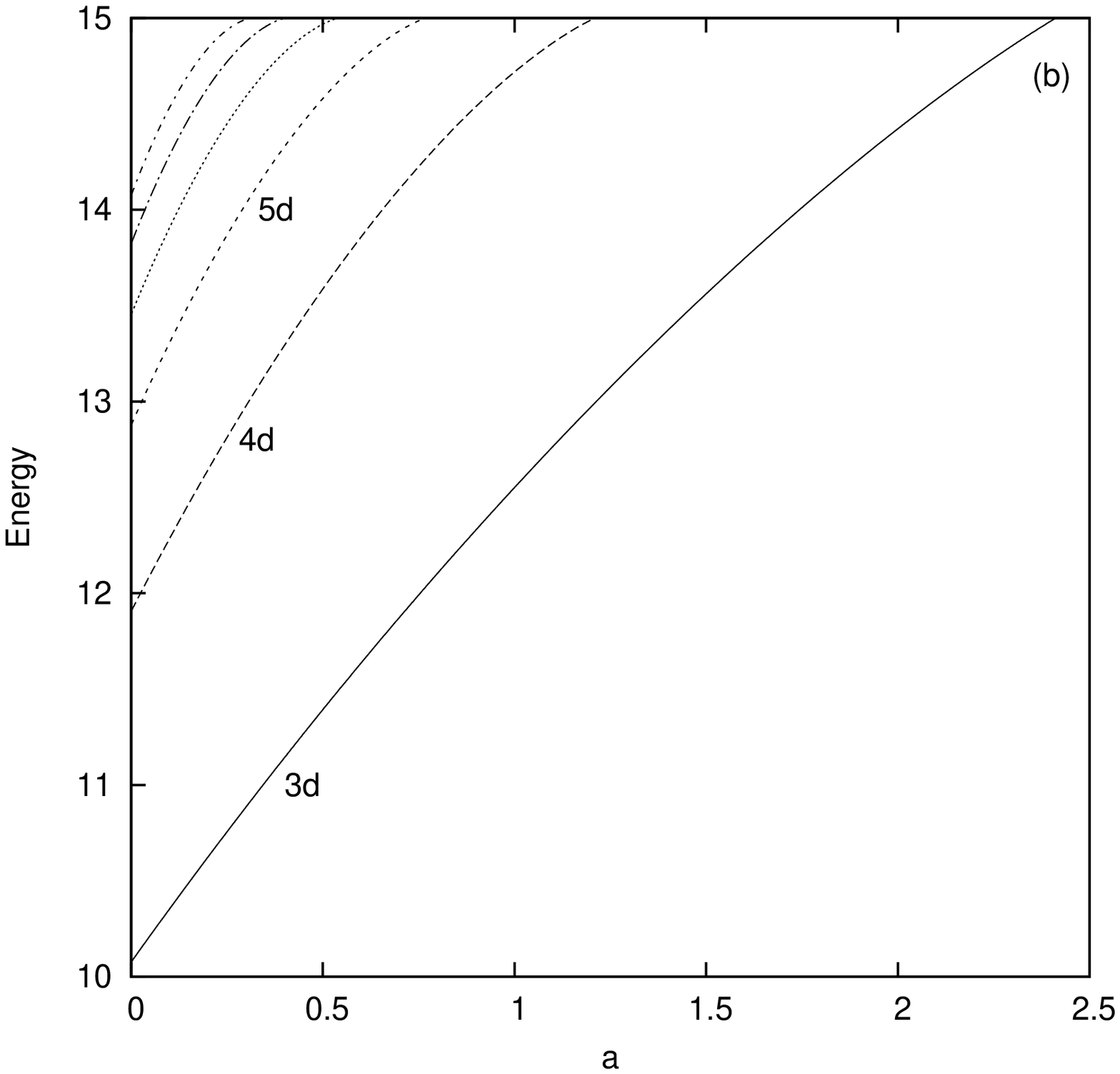}
\end{minipage}%
\hspace{0.5in}
\begin{minipage}[c]{0.40\textwidth}
\centering
\includegraphics[scale=0.38]{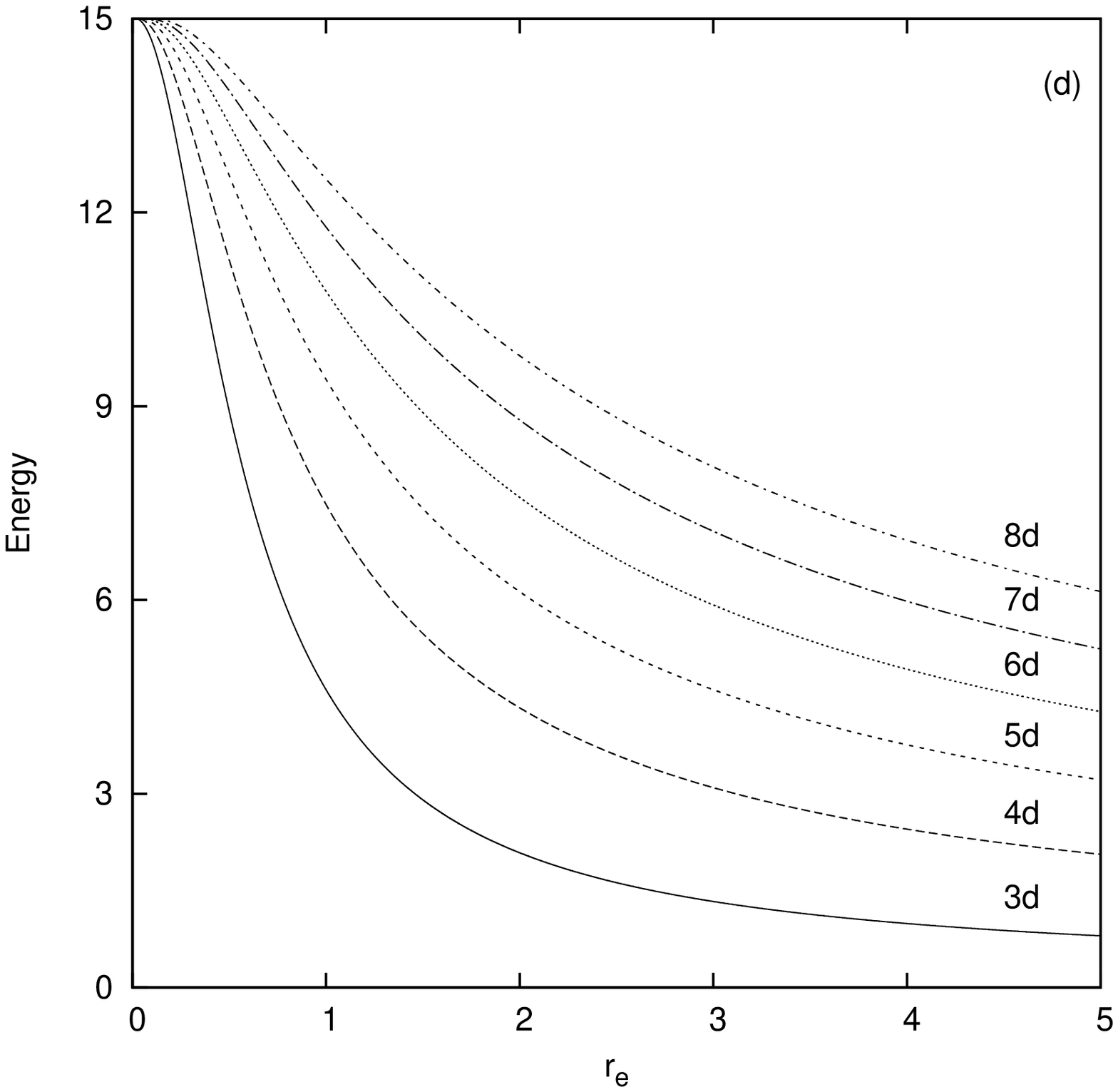}
\end{minipage}%
\\[10pt]
\begin{minipage}[b]{0.40\textwidth}\centering
\centering
\includegraphics[scale=0.38]{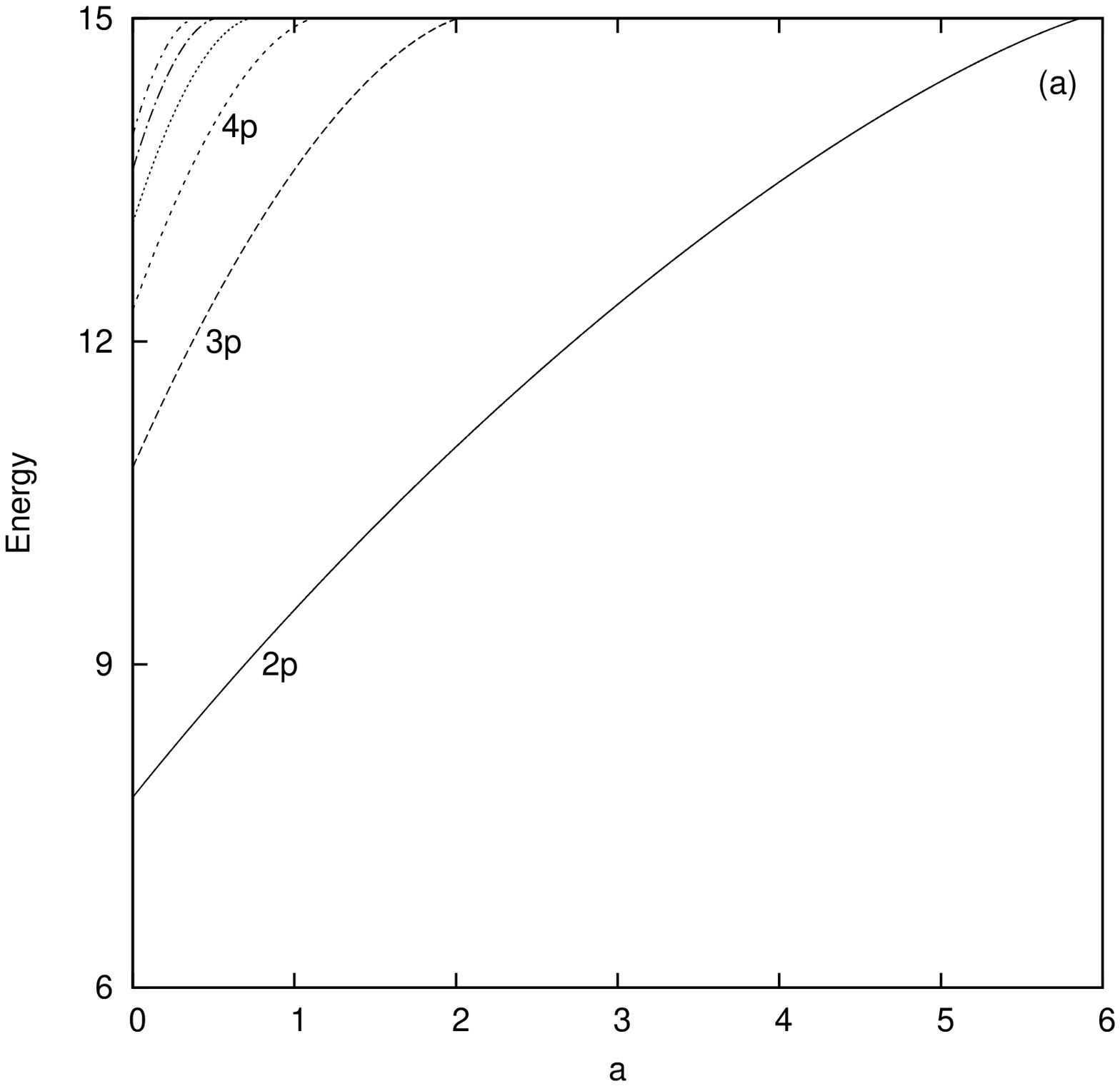}
\end{minipage}
\hspace{0.5in}
\begin{minipage}[b]{0.40\textwidth}\centering
\includegraphics[scale=0.38]{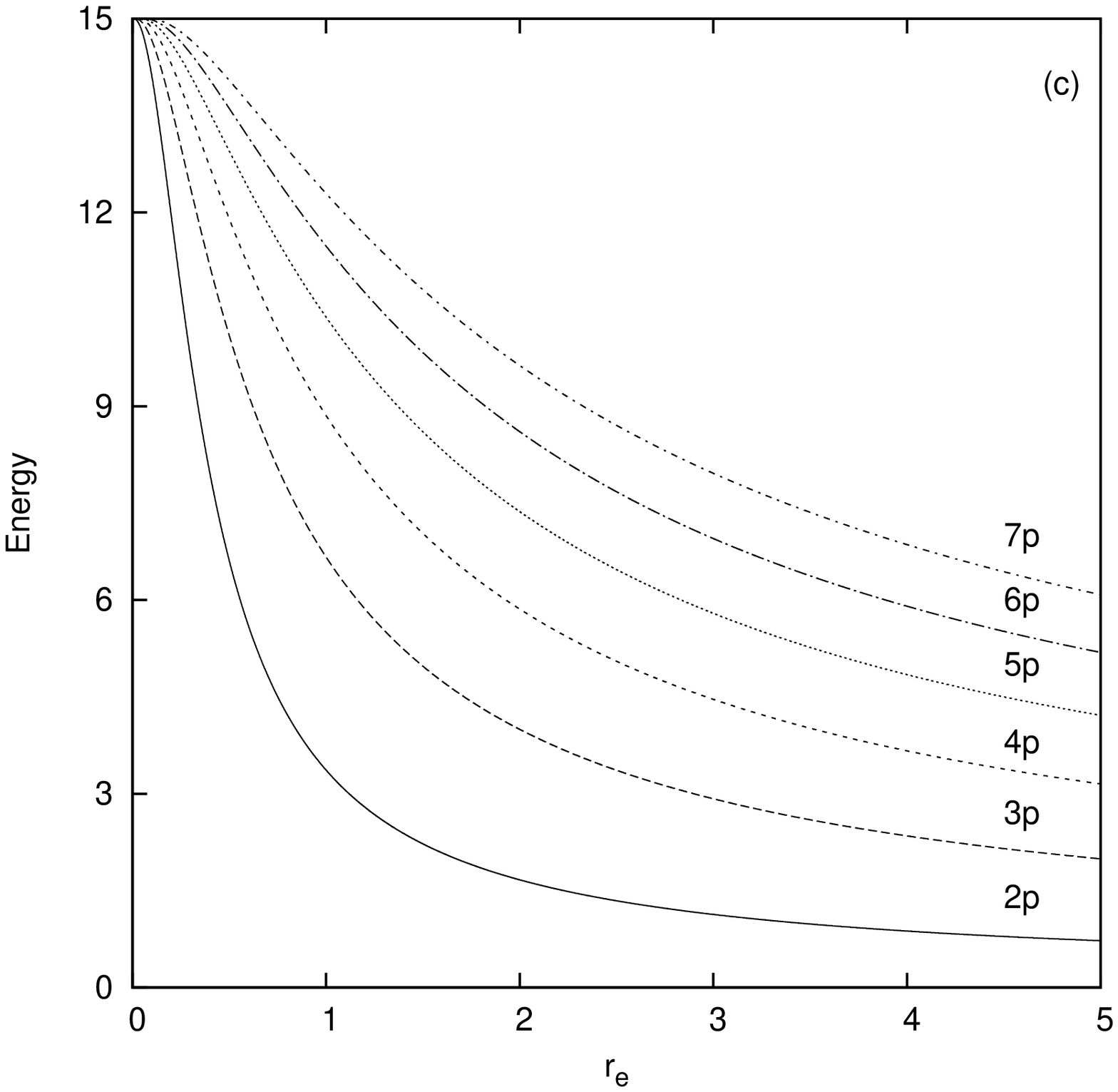}
\end{minipage}
\caption{Energy variations with respect to $a$ (left panel) and $r_e$ (right panel) of DF potential. In (a), (c), changes of $np \ (n=1-6)$ states, 
while in (b), (d), changes in $n'd \ (n'=2-7)$ states are given. The parameter $r_e$ is fixed at 0.4 in (a), (b); $a$ is fixed at 0.1 in (c), (d).}
\end{figure}

Now, we proceed for the application in molecules. Thus, Table IV presents GPS results for ro-vibrational bound states four selected molecules, 
\emph{viz.,} H$_2$, LiH, HCl and CO. These were chosen for the convenience of comparison, as maximum reference results are available for them. 
Model parameters in our calculation are directly taken from \cite{hamzavi13}; hence omitted here to avoid repetition. Conversion factors 
used in this work were taken from NIST database \cite{nist}: Bohr radius = 0.52917721092 \AA, Hartree energy = 27.21138505 eV, and electron 
rest mass = 5.48577990946 $\times 10^{-4}$ u. Calculated eigenvalues for all the nine ro-vibrational states having radial and angular 
quantum numbers $\{n,\ell\}=0,1,2$ are presented for each of these molecules. To the best of our knowledge, only ground-state energies have been 
reported before, which are referred in columns 4 and 6. It is seen that, results of \cite{hamzavi13} in the framework of Nikiforov-Uvarov 
method along with a Pekeris-type of approximation for the centrifugal term, completely coincides
with numerically obtained values from the amplitude-phase method \cite{hamzavi13}. Energies up to the tenth place of decimal have been reported 
very recently through an asymptotic iteration method \cite{oyewumi13}. Overall, the current approach offers excellent agreement with all these 
reference values, while deviations after 5--6 significant figures are encountered between present eigenvalues and those of \cite{oyewumi13}.  

\begingroup
\squeezetable
\begin{table}
\caption {\label{tab:table4}Comparison of negative eigenvalues (in eV) of $\{n,l\}=0,1,2$ states of sDF potential for H$_2$, LiH, HCl and CO, with 
literature data. PR signifies Present Result.} 
\begin{ruledtabular}
\begin{tabular}{ll|ll|ll}
$n$  &  $\ell$  & $-$E (PR)   & $-$E (Literature) &   $-$E (PR)   & $-$E (Literature) \\
\hline
    &    &   \underline{H$_2$}   &           &    \underline{LiH}     &  \\
0   & 0  & 4.39462330967   & 4.39444\footnotemark[1]$^,$\footnotemark[2],4.394619779\footnotemark[3]  & 
           2.41193395635   & 2.41195\footnotemark[1]$^,$\footnotemark[2],2.411949045\footnotemark[3]  \\
0   & 1  & 4.38018484141   &             & 2.41010642882   &       \\
0   & 2  & 4.35138602140   &             & 2.40645314111   &       \\
1   & 0  & 3.74783246520   &             & 2.21326527660   &       \\
1   & 1  & 3.73459807457   &             & 2.21151195865   &       \\
1   & 2  & 3.70820291208   &             & 2.20800705671   &       \\
2   & 0  & 3.16436589497   &             & 2.02458779362   &       \\
2   & 1  & 3.15227170121   &             & 2.02290735749   &       \\
2   & 2  & 3.12815288405   &             & 2.01954818795   &       \\
\hline
    &    &   \underline{HCl}     &            &    \underline{CO}     &  \\
0   & 0  & 4.4170494559    & 4.41705\footnotemark[1]$^,$\footnotemark[2],4.417077001\footnotemark[3]    & 
           11.0807513815   & 11.08068\footnotemark[1]$^,$\footnotemark[2],11.08075178\footnotemark[3]  \\
0   & 1  & 4.4144716335    &             & 11.0802751600   &         \\
0   & 2  & 4.4093172886    &             & 11.0793227327   &         \\
1   & 0  & 4.0282909422    &             & 10.7941665182   &         \\
1   & 1  & 4.0258070941    &             & 10.7936950552   &         \\
1   & 2  & 4.0208406973    &             & 10.7927521446   &         \\
2   & 0  & 3.6590324655    &             & 10.5116273651   &         \\
2   & 1  & 3.6566421938    &             & 10.5111606590   &         \\
2   & 2  & 3.6518629506    &             & 10.5102272623   &         \\
\end{tabular}
\end{ruledtabular}
\begin{tabbing}
$^{\mathrm{a}}$Nikiforov-Uvarov result \cite{hamzavi13}. \hspace{100pt}  \=
$^{\mathrm{b}}$Amplitude phase result \cite{hamzavi13}. \hspace{100pt}  \=
$^{\mathrm{c}}$Ref.~\cite{oyewumi13}.
\end{tabbing}
\end{table}
\endgroup

Next we move on to the higher lying states in the same four molecules. Table V tabulates 12 such eigenstates having $n=3,5,7$ for four values of 
angular quantum number $\ell=0,5,10,15$. While for $\ell =0,5,10$, some decent number of references exist, to our knowledge, no attempts have 
been made so far for $\ell$ beyond 10. The $\ell=0,5,10$ states having $n=5,7$ have been calculated before by Nikiforov-Uvarov \cite{hamzavi13}, 
amplitude \cite{hamzavi13} and asymptotic iteration methods \cite{oyewumi13}. The first two literature energies are completely identical
in all the $(n,\ell)$ states, with $n$ having 5,7 and $\ell=0$. For $\ell \neq 0$ states, however,  there remains considerable difference between 
these two references which apparently grows as $n, \ell$ increase. While for small $n, \ell$, quantum numbers our results were in quite 
good agreement with all these literature values in Table IV, there is a growing tendency of these results differing from each other for higher 
$(n,\ell)$ states. In such occasions, GPS results seem to show maximum agreement with the amplitude results \cite{hamzavi13}. Reference energies 
are unavailable for $n=3$ states. 

\begingroup
\squeezetable
\begin{table}
\caption {\label{tab:table5} Comparison of negative eigenvalues (in eV) of some high-lying states ($n=3,5,7; \ell=0,5,10,15$) of sDF potential for 
H$_2$, LiH, HCl, CO with literature data. PR signifies Present Result.} 
\begin{ruledtabular}
\begin{tabular}{ll|ll|ll}
$n$  &  $\ell$  & $-$E (PR)   & $-$E (Literature) &   $-$E (PR)   & $-$E (Literature) \\
\hline
    &    &   \underline{H$_2$}   &           &    \underline{LiH}     &  \\
3   & 0  & 2.6405317104    &         & 1.8456748599    &  \\ 
3   & 5  & 2.4776515940    &         & 1.8216012066    &   \\
3   & 10 & 2.0666963273    &         & 1.7580157392    &   \\
3   & 15 & 1.4674645655    &         & 1.6565696624    &   \\
5   & 0  & 1.7584736060    & 1.75835\footnotemark[1]$^,$\footnotemark[2],1.758451567\footnotemark[3]  & 1.5162733601    & 1.51628\footnotemark[1]$^,$\footnotemark[2],1.516277294\footnotemark[3]  \\
5   & 5  & 1.6256168674    & 1.61731\footnotemark[1],1.62548\footnotemark[2],1.617410615\footnotemark[3] & 1.4942942044    & 1.49278\footnotemark[1],1.49429\footnotemark[2],1.492771433\footnotemark[3] \\
5   & 10 & 1.2927037882    & 1.26034\footnotemark[1],1.29257\footnotemark[2],1.260451640\footnotemark[3] & 1.4362755837    & 1.43062\footnotemark[1],1.43627\footnotemark[2],1.430614300\footnotemark[3] \\
5   & 15 & 0.8150952670    &         & 1.3438205366    &   \\
7   & 0  & 1.0776596799    & 1.07756\footnotemark[1]$^,$\footnotemark[2],1.077636993\footnotemark[3]  & 1.2233927653   & 1.22340\footnotemark[1]$^,$\footnotemark[2],1.223393538\footnotemark[3] \\ 
7   & 5  & 0.9724270534    & 0.96174\footnotemark[1],0.97232\footnotemark[2],0.961814782\footnotemark[3] & 1.2034455538    & 1.20173\footnotemark[1],1.20344\footnotemark[2],1.201724343\footnotemark[3] \\
7   & 10 & 0.7118181328    & 0.66976\footnotemark[1],0.71172\footnotemark[2],0.669844065\footnotemark[3] & 1.1508305492    & 1.14444\footnotemark[1],1.15083\footnotemark[2],1.144438594\footnotemark[3] \\
7   & 15 & 0.3489528669    &         & 1.0671127231   &   \\
\hline
    &    &   \underline{HCl}     &            &    \underline{CO}     &  \\
3   & 0  & 3.3090199916    &         & 10.233121438    &  \\ 
3   & 5  & 3.2746102477    &         & 10.226192730    &                \\
3   & 10 & 3.1833288225    &         & 10.207721916    &   \\
3   & 15 & 3.0364793276    &         & 10.177724648    &   \\
5   & 0  & 2.6657422481    & 2.66574\footnotemark[1]$^,$\footnotemark[2],2.665748019\footnotemark[3]    & 9.6881596258    & 9.68809\footnotemark[1]$^,$\footnotemark[2],9.688146187\footnotemark[3]  \\
5   & 5  & 2.6341202067    & 2.62859\footnotemark[1],2.63411\footnotemark[2],2.628601192\footnotemark[3] & 9.6813735596    & 9.68017\footnotemark[1],9.68130\footnotemark[2],9.680226284\footnotemark[3] \\
5   & 10 & 2.5502777586    & 2.52989\footnotemark[1]2.55027\footnotemark[2],2.529905688\footnotemark[3] & 9.6632831420    & 9.65905\footnotemark[1],9.66321\footnotemark[2],9.659110919\footnotemark[3] \\
5   & 15 & 2.4155342071    &         & 9.6339040805    &   \\
7   & 0  & 2.0965250897    & 2.09652\footnotemark[1]$^,$\footnotemark[2],2.096524802\footnotemark[3]  & 9.1591824044    & 9.15911\footnotemark[1]$^,$\footnotemark[2],9.159164003\footnotemark[3]  \\ 
7   & 5  & 2.0676862795    & 2.06161\footnotemark[1],2.06768\footnotemark[2],2.061620020\footnotemark[3] & 9.1525389621    & 9.15131\footnotemark[1],9.15247\footnotemark[2],9.151359661\footnotemark[3] \\
7   & 10 & 1.9912752181    & 1.96888\footnotemark[1],1.99127\footnotemark[2],1.968892038\footnotemark[3] & 9.1348288985    & 9.13050\footnotemark[1],9.13476\footnotemark[2],9.130552425\footnotemark[3] \\
7   & 15 & 1.8686394018    &         & 9.1060679852    &   \\
\end{tabular}
\end{ruledtabular}
\begin{tabbing}
$^{\mathrm{a}}$Nikiforov-Uvarov result \cite{hamzavi13}. \hspace{100pt}  \=
$^{\mathrm{b}}$Amplitude phase result \cite{hamzavi13}. \hspace{100pt}  \=
$^{\mathrm{c}}$Ref.~\cite{oyewumi13}.
\end{tabbing}
\end{table}
\endgroup

The above variations in energy for molecules are depicted in Fig.~2. Representative plots are given for two molecules, namely, H$_2$ and LiH 
in the lower and upper segments respectively. The lower (a) and upper (b) panels on the left hand side correspond to such changes in energy as the
vibrational quantum number $n$ varies, for H$_2$ and LiH respectively. In both cases, six values of $\ell$ are chosen, i.e., $0,5,10,15,20,25$. 
Note that in H$_2$, the $n$ scale goes to 15, whereas for LiH, the same is shown up to 25. This is because of the fact that, this potential supports 
a limited number of bound states only; these are available in lesser number in H$_2$ than in LiH. The bottom (c) and top (d) segments on the 
right side, likewise, show energy variations with respect to angular quantum number $\ell$. These are given for five $(0,3,6,9,12)$ and 
six $(0,3,6,9,12,15)$ values of $n$, for H$_2$ and LiH respectively. For a given molecule, $E_{n,\ell}$ versus $\ell$ tends to attain a straight 
line-like behavior for higher and higher $n$; similarly $E_{n,\ell}$ versus $n$ also approaches a linear behavior for progressively higher $\ell$. 
In moving from H$_2$ to LiH, the $E_{n,\ell}$ versus $n$ tends to become more closely spaced, whereas the $E_{n,\ell}$ versus $\ell$ remain well 
distinctly separated, although the rate of change slows down bringing some flatness in to the picture. The general qualitative features of these 
plots remain quite similar for the other two molecules, HCl and CO. $E_{n\ell}$ versus $n$ in HCl remains very close to
that of LiH, while for CO, the individual $\ell$ plots become \emph{much closer} to each other and assuming almost linear behavior, much like 
the way in \cite{roy13a} for Morse potential. The $E_{n,\ell}$ 
versus $\ell$ plot of HCl, once again resembles very closely that of LiH, while in CO, the individual $n$ plots remain well separated however.
These energy variations show good resemblance with those obtained in the recent study of Morse potentials for same set of molecules \cite{roy13a}. 
This is in good accord with the recent finding of \cite{wang12}, where the anharmonicity $\omega_e \chi_e$ and vibrational rotational coupling 
parameter $\alpha_e$ for 16 selected molecules were found to be quite similar for the DF and Morse potential. 
To the best of our knowledge, no such analysis has been made before. We hope that the present results may provide useful guidelines for future works. 

\begin{figure}
\centering
\begin{minipage}[t]{0.40\textwidth}\centering
\includegraphics[scale=0.38]{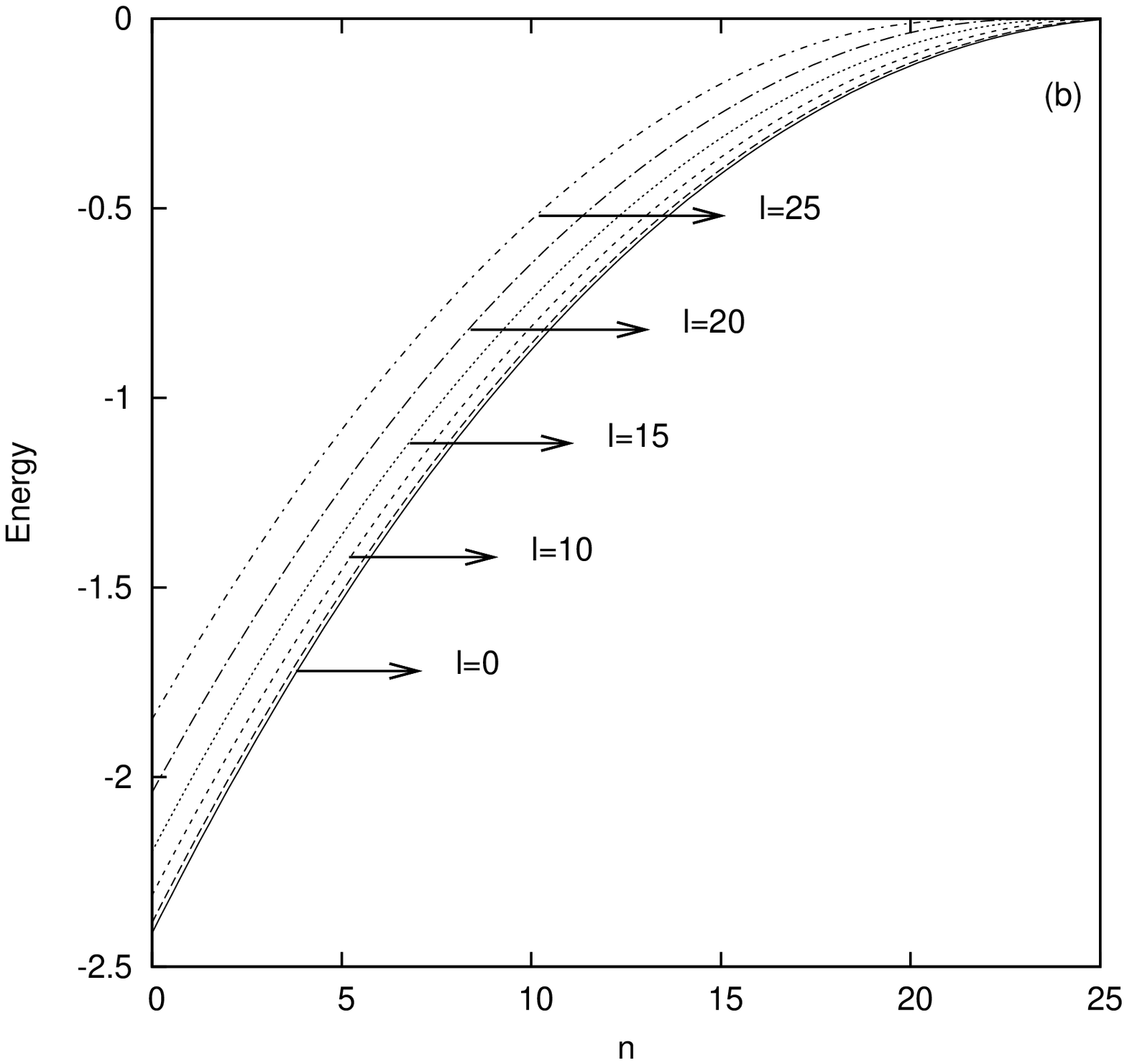}
\end{minipage}
\hspace{0.15in}
\begin{minipage}[t]{0.35\textwidth}\centering
\includegraphics[scale=0.38]{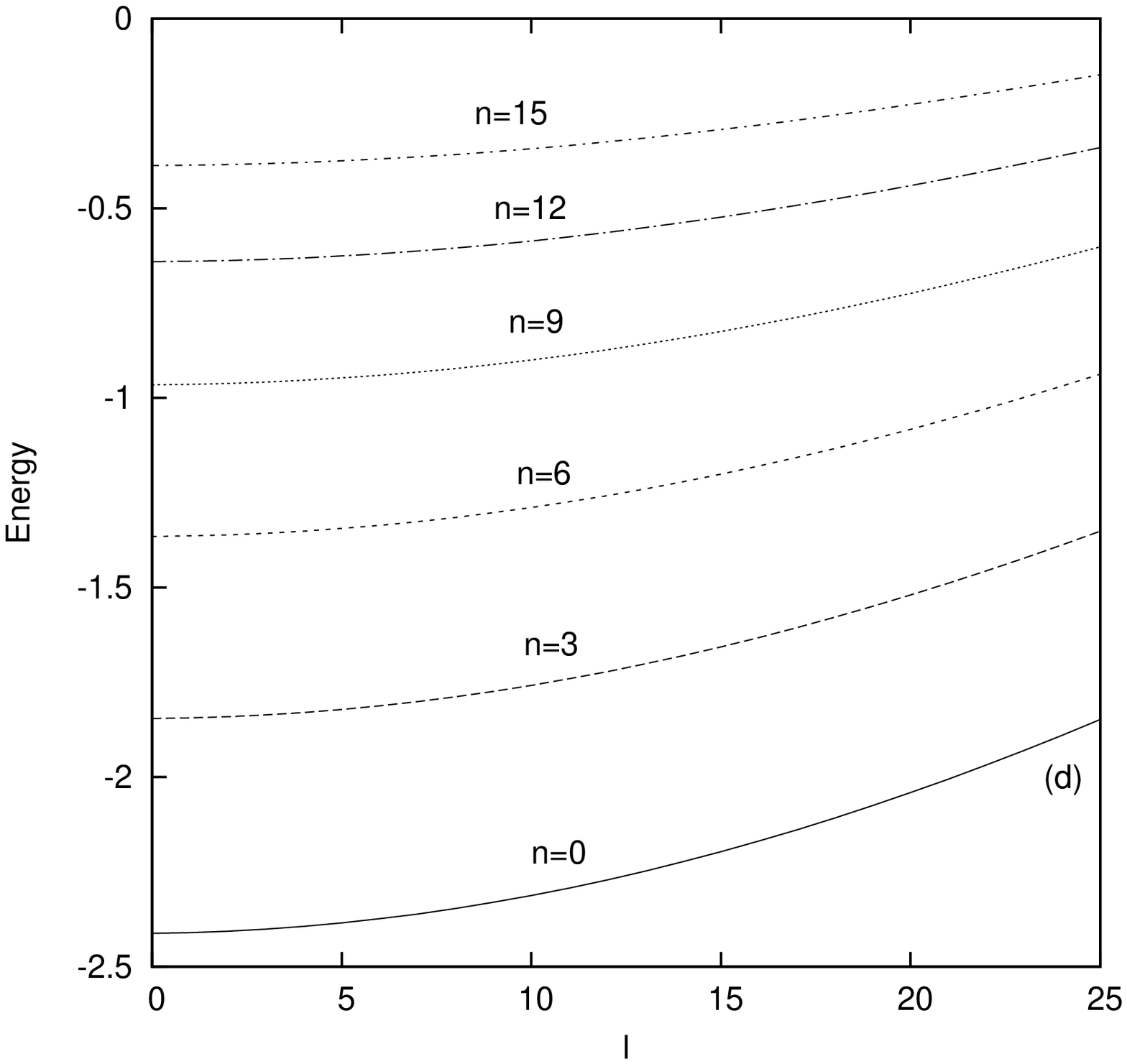}
\end{minipage}
\\[10pt]
\begin{minipage}[b]{0.40\textwidth}\centering
\includegraphics[scale=0.38]{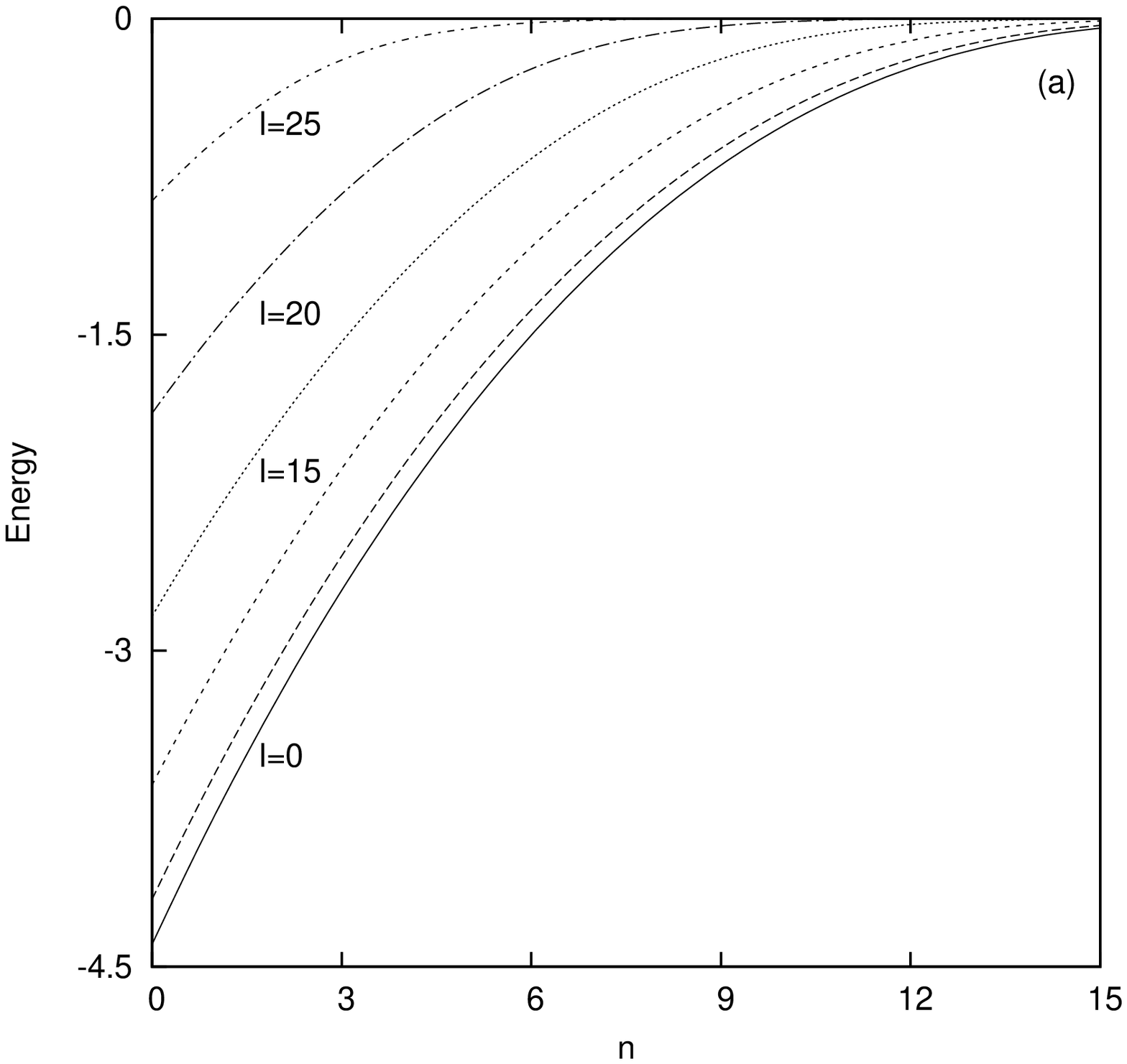}
\end{minipage}
\hspace{0.15in}
\begin{minipage}[b]{0.35\textwidth}\centering
\includegraphics[scale=0.38]{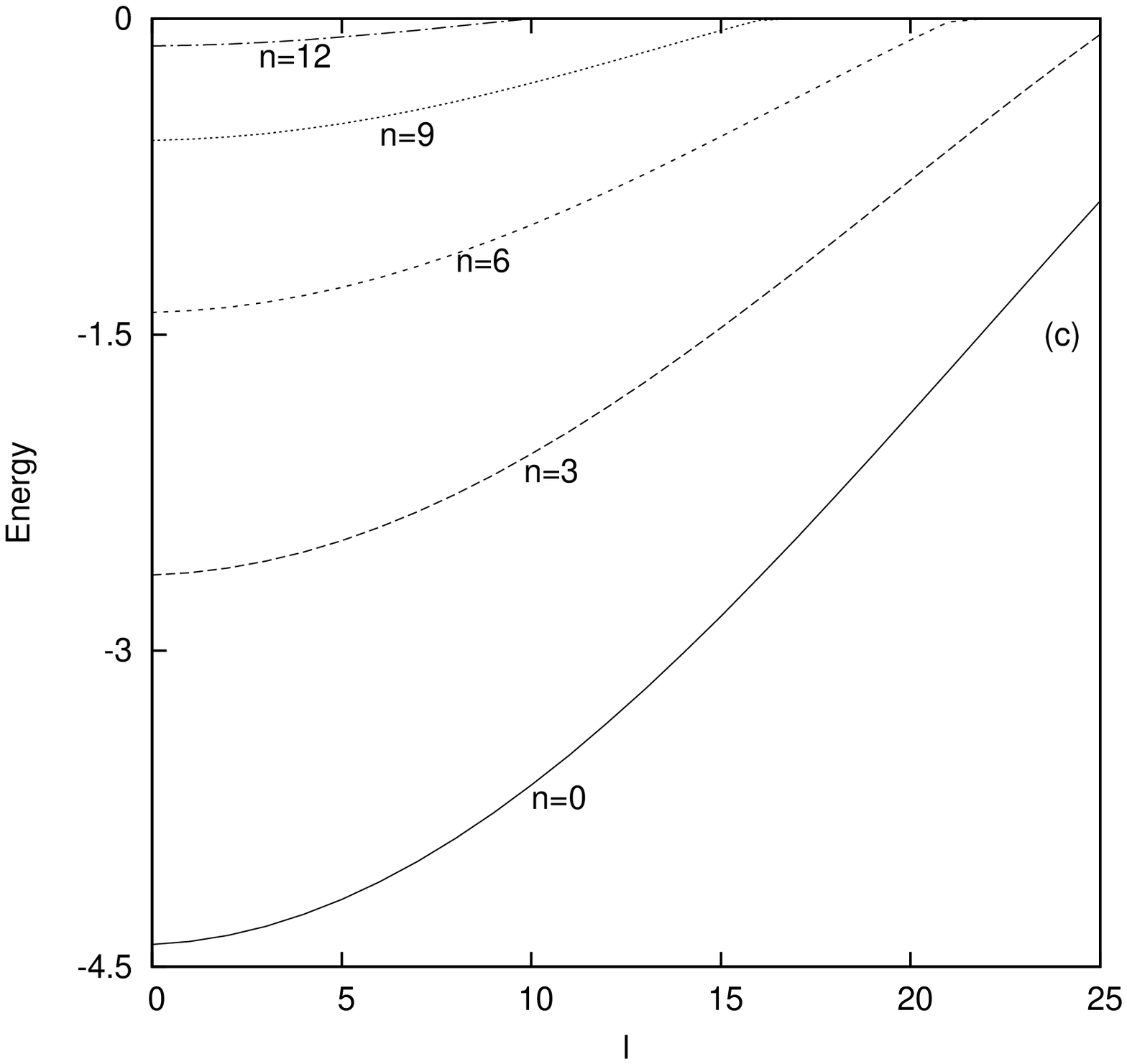}
\end{minipage}
\caption[optional]{Energy variations (in eV) of sDF potential, with respect to vibrational ($n$) and rotational ($\ell$) quantum numbers in left 
and right panel respectively. In the former, six $\ell$ values, \emph{viz.,} $0,5,10,15,20,25,$ while for the latter, five $(0,3,6,9,12)$ and 
six $(0,3,6,9,12,15)$ $n$ values were selected. Lower and upper portions correspond to H$_2$, LiH respectively.}
\end{figure}

\section{conclusion}
Accurate bound-state energies of DF and sDF potential are presented within a GPS formalism. Low and high eigenstates are calculated with very good 
accuracy, as demonstrated by comparing with the literature data. Excellent agreement has been observed in all cases. First, 21 states belonging 
to radial and angular quantum numbers up to 5 are studied. Then, states belonging to $\{n,\ell\}$ quantum numbers having values greater than 5 
(15 states with $n=6,7$) are reported here, for the first time. Energies are calculated over a large range of the potential parameters $r_e, D_e$
to analyze their dependence. Then this is applied for the vibration-rotation of four diatomic molecules, which shows very good agreement with
existing results. Furthermore, a close examination of the energy changes with respect to state indices $n$, $\ell$ reveals very similar behavior
with those offered by the traditional Morse potential. The GPS methodology is simple, easy to implement and efficient. Yet this offers results 
comparable to other more complicated methods. Given its success for this and previous systems, we hope the method will be equally applicable to 
other relevant potentials in atomic and molecular physics.

\section{acknowledgment} It is a pleasure to thank the IISER-Kolkata colleagues for many fruitful discussions. I am grateful to the Director, 
Prof.~R.~N.~Mukherjee, for his kind support. I sincerely thank the two anonymous referees for their kind, constructive and valuable comments, from which 
the manuscript has greatly benefited.

\end{document}